\documentstyle[12pt,aasms4,tighten]{article}

\newcommand{\kms}{$\rm km \, s^{-1}$}
\def\etal{{\it et~al.\ }}
\def\eg{{\it e.g.\ }}
\def\ie{{\it i.e.,\ }}

\begin{document}
  
\title{The Parker Instability in 3-D: Corrugations and Superclouds Along 
the Carina-Sagittarius Arm} \author{Jos\'e Franco\altaffilmark{1}, 
Jongsoo Kim\altaffilmark{2}, Emilio J. Alfaro\altaffilmark{3} and Seung 
Soo Hong\altaffilmark{4}}

\altaffiltext{1}
{Instituto de Astronom\'{\i}a, Universidad Nacional Aut\'onoma de
M\'exico, Apdo. Postal 70-264, 04510 M\'exico D. F., M\'exico;
pepe@astroscu.unam.mx}
\altaffiltext{2}
{Korea Astronomy Observatory, 61-1, Hwaam-Dong, Yusong-Ku, Taejon 305-348,
Korea; jskim@kao.re.kr}
\altaffiltext{3}
{Instituto de Astrof\'{\i}sica de Andaluc\'{\i}a CSIC, Apdo. 3004, Granada
18080, Spain; emilio@iaa.es}
\altaffiltext{4}
{Department of Astronomy, Seoul National University, Seoul 151-742, Korea;
sshong@astroism.snu.ac.kr}

\begin{abstract}
Here we present three-dimensional MHD models for the Parker instability
in a thick magnetized disk, including the presence of a spiral arm. The
$B$-field is assumed parallel to the arm, and the model results are applied
to the optical segment of the Carina-Sagittarius arm. The characteristic
features of the
undular and interchange modes are clearly apparent in the simulations. The
interchange mode appears first and generates small interstellar structures
in the inter-arm regions, but its development inside the arm is hampered
by the acceleration of the spiral wave. In contrast, the undular mode
follows its normal evolution inside the spiral wave, creating large gas
concentrations distributed along the arm. This results in a clear
arm/inter-arm difference: the instability triggers the formation of large
interstellar clouds (with masses in the range of $10^6$ to $10^7$ $M_{\odot}$)
inside the arms, but generates only small structures with slight density
enhancements in the inter-arm regions. The resulting clouds are distributed in
an antisymmetric way with respect to the midplane, and their masses are similar
to those inferred for HI superclouds in our Galaxy. Such a cloud distribution
results in an azimuthal corrugation along the arm and, for conditions similar to
those of the optical segment of the Carina-Sagittarius arm, it has a wavelength
of about 2.4 kpc. This structuring, then, can explain the origin of both HI
superclouds and the azimuthal corrugations in spiral arms. In particular, the
wavelength of the fastest growing undular mode matches the corrugation length
derived with the young stellar groups located in the optical segment of the
Carina-Sagittarius arm.
\end{abstract}


\section{INTRODUCTION} 

The gaseous disk of the Milky Way is a magnetized medium with a variety of
different components, some of them extending for a few hundred parsecs above
the midplane (see Boulares \& Cox 1990). It has a very complex structure, 
with atomic and molecular clouds threaded by magnetic fields and embedded in
a more diffuse and ionized medium. Also, the gas of the disk is continuously
perturbed by a number of different agents. These range from well defined and
localized energy perturbations, such as expanding HII regions and supernova 
explosions, to large scale shocks driven by spiral waves (or by global 
perturbations induced by interactions with nearby neighbors). The magnetic 
field provides partial support to the extended disk layers (and allows for an
efficient energy and momentum exchange between different parts of the disk), 
but also induces large scale instabilities, as discussed several decades 
ago by Parker (1966, 1967). When a relatively well ordered $B$-field is 
present, the compression generated by a strong perturbation (say, a spiral 
wave) changes the magnetic field downstream and the gas driven by the distorted
field lines accumulates large mass clouds along the valleys in the field lines.
This is called the ``Parker'' instability and, as in the case of gravitational 
instabilities, it may also gather giant cloud complexes in spiral arms (\eg
Mouschovias \etal 1974; Shu 1974; Elmegreen \& Elmegreen 1986). Obviously, these
type of instabilities can be triggered by a variety of different perturbations 
(\eg Franco \etal 1995; Hanasz \& Lesch 2000), and a two-dimensional (2-D) 
analysis of the terms driving separate instabilities in a sheared disk (\ie 
thermal, gravitational, and Parker) has been reviewed by Elmegreen (1992,1993).
More recently, with a 3-D model of the Parker instability in a thin disk and 
under a uniform gravity, the actual possibility of forming large cloud 
complexes via this instability has been questioned (\eg Kim \etal 1998), but 
a more detailed study with a realistic galactic disk model in 3-D is now 
needed. This is indeed important because our Galaxy has a thick, 
multi-component gaseous disk with a relatively strong magnetic field that 
extends more than a kiloparsec above the plane of the disk(\eg Boulares \&
Cox 1990; Beck et al. 1996). In contrast with previous thin disk models, the 
presence of these extended magnetized components change the response of the 
general interstellar medium to perturbations (\eg Martos \& Cox 1998; Martos 
\etal 1999; Santill\'an \etal 1999).

The first steps in this direction, including a multicomponent model for the
gaseous disk and the observed gravitational field in the solar neighborhood, 
have been done by Kim \etal (2000; hereafter Paper I) and Santill\'an \etal 
(2000; hereafter Paper II). They performed a 2-D linear stability analysis, 
with the corresponding MHD models, and found that the wavelengths and time 
scales for the Parker instability are strongly modified. The most unstable 
undular mode in the Solar circle has a wavelength of about 3 kpc and an 
e-folding time of $\sim 3\times 10^7$ yr (significantly larger than those 
derived for thin disk cases; see Kim \& Hong 1998). The 2-D MHD experiments 
confirm these results and show, in addition, that the preferred parity of the 
instability is antisymmetrical (\ie the resulting condensations are distributed
above and below the midplane). Thus, one signature of the undular mode is a 
corrugation pattern along the direction of the original $B$-field. 

The existence of corrugations along spiral arms has been discussed in the 
literature by several authors, and perhaps the best studied case is the 
Carina-Sagittarius arm. There are several important problems in deriving the 
precise location and properties of spiral arms in the Milky Way (see discussions
by Liszt 1985 and Roberts \& Burton 1997), and most properties of spiral 
structure are better studied in external galaxies. Nonetheless, even when its
detailed properties are not completely established, the Carina-Sagittarius arm 
is one of the most conspicuous and best studied spiral arms in our Galaxy. 
According to present estimates, it circles for about 40 kpc around the galactic 
center, and follows a logarithmic spiral curve with a pitch angle of $\sim 
10^{\circ}$, close to the one expected from the density wave theory (\eg 
Grabelsky \etal 1988). The wavy vertical distribution of large clouds along 
Carina-Sagittarius unveils an azimuthal corrugation pattern that seems to be 
present in other arms of our Galaxy (Spicker \& Feitzinger 1986). Such a pattern
is also apparent in the distribution of young stellar clusters within the arm 
(Alfaro \etal 1992a). In contrast, the existence of azimuthal wavy structuring 
has not been reported for the inter-arm regions. It is unclear, however, if this
is simply due to observational restrictions (\ie due to a smaller amount of 
gaseous features and young stellar tracers) or to a real lack of structuring. 

With all these restrictions in mind, here we explore the features that appear 
in 3-D models for the Parker instability operating inside the arms. 
The instability has two independent modes, undular and interchange, with 
different properties and wavelengths (see Hughes \& Cattaneo 1987). The undular
mode, which is the best studied branch of the instability, is a 2-D event that 
evolves in the plane defined by the gravitational acceleration and the original
direction of the $B$-field. It deforms the magnetic field lines, and creates 
large gas condensations distributed in an antisymmetric fashion with respect to
the midplane. Such an arrangement gives the impression of a corrugated system, 
with a corrugation length equal to the wavelength of the fastest growing mode.
The interchange mode, on the other hand, operates outside the plane of the
undular mode and allows for the vertical exchange of field lines. In 3-D, both 
modes combine into a mixed mode, creating a more complex network of
condensations. However, some of the features generated by the pure undular
mode can persist in the non-linear 3-D regime, but the resulting gas
structures are more filamentary (Kim \etal 1998; Kim, Ryu \& Jones 2001).
Thus, as long as the undular structure is not washed away by the action of
the interchange mode in 3-D, the structuring observed along spiral arms may well
be explained by this instability. As pointed 
out by Beck \etal (1989), this is also a potential mechanism to create the 
periodic variations observed in the $B$-field of the southwestern arm of M31. 

In this paper we show that this is indeed the case, and report the results of 
3-D magneto-hydrodynamical simulations of the Parker instability in a thick 
disk including the presence of a spiral wave. The model is
performed with parameters appropriate for the optical segment of
Carina-Sagittarius, the best studied region of this arm. The presence
of the spiral wave is modeled as a perturbation in the gravitational
field. The interchange mode appears at early times but the acceleration of
the spiral wave, even when small compared with the disk gravity,
effectively quenches its further development inside the arm. Hence, the
undular mode ends up dominating the evolution inside the spiral wave, and
creates major gas condensations along the arm. These condensations are
distributed in an antisymmetric way with respect to the midplane, and with
a wavelength of about 2.4 kpc, similar to the corrugation length derived 
by Alfaro \etal (1992a) and Berdnikov \& Efremov (1993). The organization 
of the paper is as follows. Section 2 describes the 3-D models and results. 
Section 3 presents an overview of the observed properties of the 
Carina-Sagittarius Arm, and Section 4 presents a brief discussion of the 
results.

\section{Three--Dimensional Models}

\subsection{The Model Galaxy}

The details of the multi-component thick disk model for the Solar circle (\ie 
the observed disk gravity and gas density stratifications near the Sun) are 
described in Papers I and II. Here we use the same model but scale the disk 
parameters to those of the optical segment of Carina-Sagittarius (see next 
Section). This is perhaps the best known region of the arm
and is located inside the solar circle at galactocentric radii between 
$R/R_{0}\sim$ 0.8 -- 0.9, where $R_{0} \sim 8.5$ kpc is the radius of the Solar
circle. As stated before, Carina-Sagittarius has a total extension of about 40 
kpc, and covers a wide range of galactocentric radii, from about 5 to 12 kpc 
(these values are corrected for the location of the Sun; see Fig. 4 in 
Grabelsky \etal 1988). Given that the scale height and midplane density of the 
gaseous disk vary with galactocentric distance, the average value of these 
parameters change by about 40\% along the radial extent of this arm (see 
Malhotra 1994). Using the radial distributions of molecular and atomic gas in 
our Galaxy reported by Malhotra (1994, 1995), the midplane density at the 
location of the optical segment increases by 20-30 \% with respect to the Solar
circle, whereas the HI scale height decreases by a similar amount. Also, the 
gravitational acceleration of the disk at these locations is increased by 25 \%.

A brief summary of the magnetic field in our Galaxy is given in the next section,
and here we assume a well ordered field, with a midplane strength of $B(0)=5$ 
$\mu$G, that runs parallel to the spiral arm (see Beck \etal 1996; Heiles 
1996; Indrani \& 
Deshpande 1998; Vallee 1998). The spiral density wave representing the arm is 
modeled as a perturbation in the gravitational field, following the elliptical 
approximation described by Martos \& Cox (1998). This perturbation, which has a
maximum amplitude of 5\% of the gravitational field of the disk, is an ellipse 
centered in the $x-z$ plane, and goes to zero in a distance of 1 kpc along the 
$x$-direction and 0.6 kpc along the $z$-direction.

In summary, we use the $z$-distributions described in Papers I and II with
some modifications: the gas density is increased by 25 \% ($n_0=1.4$
cm$^{-3}$), the effective scale height is decreased by 20 \% ($H_{\rm eff}=
133$ pc), and the gravitational acceleration is increased by 25 \%. An
additional 5 \% gravitational perturbation is then added at the location of
the arm. All gas layers are assumed isothermal, with a sound speed of $c_s=8.4$
\kms. The basic time unit is $H_{\rm eff}/c_s\sim 1.54 \times 10^7$ yr and,
unless stated otherwise, all times are given in this unit. This disk is Parker
unstable and the resulting most unstable perturbation for the undular mode,
derived with the method described in Paper I, has a wavelength of $\sim 2.4$ 
kpc and its e-folding time is $\sim 3 \times 10^7$ yr.

The models are performed with the isothermal MHD TVD code described by 
Kim \etal (1999) in a Cartesian coordinate system, $(x,y,z)$, with a 
128$^3$ zone grid. The frame of reference is at rest with the spiral wave
and, for simplicity, the gas has no relative motion with respect to the 
spiral arm. The physical size of the computational cube is (4 kpc)$^3$, 
and the linear resolution of the models is about 32 pc. The spiral arm is
oriented along the $y$-axis, and the $z$-axis is perpendicular to the 
midplane (the initial $B$-field direction, then, is also along the 
$y$-axis). Galactic differential rotation and self-gravity are not 
included, and we restrict the study to the structures formed during time
scales of the order of several times $10^8$ yr.

For the initial setup, as described in Paper II, the unperturbed disk is
set in magnetohydrostatic equilibrium at $t=0$. Then, the acceleration of
the spiral wave is introduced in the $x$ and $z$ directions of the
computational domain and, to generate a perturbation in the $y$-axis, an
additional velocity perturbation is also introduced at this moment. We
made a series of runs with different types of perturbations, including
random and sinusoidal velocity fields, and with several initial amplitudes
(ranging from $10^{-4}$ to $10^{-1}$ $c_s$). The spiral wave and the
velocity perturbations, then, represent the seeds that trigger the
instability. 

\subsection{Results}

Figure 1 shows the evolution of the rms gas velocity
components ($<v_x^2>^{1/2}$, $<v_y^2>^{1/2}$, and $<v_z^2>^{1/2}$) for the
run with a random velocity perturbation and with an initial amplitude of
$10^{-4}\ c_s$. The solid line has a slope equal to the growth rate of the
fastest growing undular mode. The gas motions along the $y$-direction are
not affected by the gravitational acceleration of the spiral wave and,
except for a short period at very early times (\ie $t \leq 10$), the
evolution of $v_y$ follows the track expected for the undular mode (see
Paper II). In contrast, the motions along the other two directions are
continuously modified by the acceleration of the arm, and the evolution of
$v_x$ and $v_z$ follow different growth rates up to $t\sim 20$. At about
$t=25$, all three velocity components reach a maximum value and then decline
with smooth oscillations at later times. Thus, the evolution can be divided
into the three stages of the instability (linear, non-linear, and final
equilibrium) described in Paper II. The initial linear stage lasts up to
about $t=20$, and the non-linear saturated stage can be defined up to
about $t\sim 40$. The system enters into the final equilibrium (or damping
oscillatory) stage after this time.
These same features also appear in the rest of the runs with sinusoidal
perturbations, and they only differ in the timescales at which the different
stages are initiated (for sinusoidal perturbations with larger initial 
amplitudes, the linear stage starts and gets saturated at earlier times).

Figure 2 shows snapshots of the density structures, velocity fields, and
magnetic field topologies at three selected times. To simplify
the visualization of the 3-D data cubes, the densities and velocities are
shown for the central $x-z$ plane only, and the $B$-field lines are shown
only near the central $y-z$ plane. In addition, to provide a better idea of
the 3-D structuring, two isodensity surfaces (with the same reference
value $n=0.04$ cm$^{-3}$) are also included in the snapshots. The velocity fields
are represented with red arrows (the value of the unit arrow is rescaled
at each time frame), and the $B$-field lines are shown in blue. The
densities are color-coded from red to purple, as the value decreases. To
complete the information, Figure 3 shows the corresponding $x-y$ maps of
the gas column density, integrated along the $z$-axis, at the same
three selected times. The vertical column densities at these three 
epochs are normalized with the initial column density, $1.1 \times 
10^{21}$ cm$^{-2}$, which is obtained by integrating the initial density 
distribution from -2 kpc to +2 kpc (see, equation~5 in Paper II).  As
explained in the previous section, we have adjusted the midplane density
and scale height of the gas to the values in the Carina-Sagittarius region.
The resulting total column density, however, is the same as that of 
the solar neighborhood.

Once the spiral and random perturbations are introduced, the initial
magnetohydrostatic equilibrium is broken and the system begins to
oscillate. The first frames of Figs. 2 and 3, at $t=15$, illustrate the
evolution at the early stages of the instability, during the linear
growth. Given that the volume and column densities are increased by the
acceleration of the spiral wave, the location of the arm is clearly seen
in both frames. The $B$-field shows small undulations, the isodensity
surfaces are only mildly distorted, and the gas velocities at high
latitudes are being amplified. The second frames, at $t=25$, show the
evolution near the end of the linear stage. The $B$-field lines are now 
clearly twisted and strongly undulated by the combined action of the
undular and interchange modes (this is, by the mixed mode). The
interchange mode, which has the shorter wavelengths and fastest growing
rates, creates small structures during the early stages, and the isodensity
surfaces are completely deformed by the presence of several small clumps.
These are best seen in the column density map of Fig. 3, which shows that
the early structures are in general small, but the smaller ones appear
mainly in the inter-arm regions. This in turn indicates that the arm
acceleration along the $x$-direction effectively quenches the further growth
of the interchange mode (which otherwise would be slicing the computational
domain into thin filamentary structures). The effect appears in models
including the self-gravity of the gas (\eg Lee \& Hong 2001). Hence, at
later times, the regions inside the spiral arm are mainly subject to the
undular mode, but the inter-arm regions follow the mixed mode with the
combined undular and interchange branches. This is illustrated in the third
frames in Figs. 2 and 3, at $t=32$, with the evolution at the non-linear
saturated stage. There are four main condensations located along the arm,
and separated by distances between 1 and 1.3 kpc. They are positioned at
the valleys of the wavy $B$-field lines, indicating that the undular mode
is dominating the evolution along the arm. As time passes by, the condensations
continue to accumulate mass, increasing their sizes and peak densities, but
the results for times larger than $t \sim 30$ (which corresponds to $\sim 5
\times 10^8$ yr) are probably not relevant for actual applications to our
Galaxy. From the third frame of Fig. 3, then, we estimate the masses of
the largest and smallest condensations (centered at $x=0.0$ kpc and $y=-0.4$
kpc, and $x=$0.2 kpc and $y=$0.4 kpc, respectively) by integrating inside
column densities that are 1.5 times above the initial value,
$1.1 \times 10^{21}$ cm$^{-2}$.  Assuming a 10\% number ratio of helium to 
hydrogen, the resulting
masses are $2.9 \times 10^6$ $M_{\odot}$ for the small condensation, and
$9.0 \times 10^6$ $M_{\odot}$ for the large one.

Figure 4a shows the mid-plane density map at $t=32$, along with a cubic 
polynomial fitting curve that passes through the main density peaks. The map 
shows the same features that appear in the third frame of Fig. 3, indicating 
that the structuring in the column density is mainly determined by the 
structuring near the midplane. In order to see better the vertical 
distribution of these density peaks, Fig. 4b shows the vertical distribution 
of the density along the fitting curve. This panel clearly shows the 
corrugation pattern, where the density peaks are positioned alternatively 
above and below the mid-plane. Figures 5a and 5b show the magnetic
field structure near the central $y-z$ plane and near the central $x-y$ plane,
respectively. Also, we show the isodensity surfaces with 0.8 times the 
initial value, which provide a clear visual impression of the relative location
of the main gas concentrations.

Finally, Figure 6 shows the $z$-velocity component of the gas concentrations
along the arm. The Figure displays a position-velocity map for the arm at 
$t=32$, as seen from the north galactic pole. This corresponds to the expected 
gas velocity structure that may appear along the spiral arms in face-on 
galaxies. Given the restrictions of the present 3-D simulations in a Cartesian 
grid (\ie a single fluid at rest with the arm and without self-gravity, 
differential rotation, random $B$-fields, etc.), we cannot properly address the 
details of the gas motions in the $x-y$ plane and restrict the kinematic
information of the model to the $z$-velocity components of the condensations. 
Thus, this figure provides an important observable signature of the process, 
and represemts one of the main possible links between CO and HI observations 
and the present stage of the simulations. 

\section{Corrugations, Spacings, and Superclouds}

\subsection{The Carina-Sagittarius Arm}

The evidence that Carina-Sagittarius is truly a major spiral arm was discussed 
by Cohen \etal (1985) and Grabelsky \etal (1988), who delineated its location 
from the distribution of molecular clouds in the first and fourth galactic 
quadrants. Additional studies with different galactic tracers (\ie including HI
clouds, HII regions, and young stellar objects; see Alfaro \etal 1992a,b and 
references therein) support the idea that the optical segments of Carina and 
Sagittarius are indeed part of the same major spiral arm. Thus, this arm is our
best local probe for the spiral structure of the Milky Way, and for the 
distribution and spacing of gas clouds (and star forming centers) along the 
arms.

The azimuthal corrugations along this arm have been observed in both the 
stellar and gas components. The corrugation pattern reported by Alfaro \etal
(1992a) is outlined by open clusters located in the optical segment of the arm
(the closest region to the Sun, at $R/R_{0}\sim$ 0.8 -- 0.9). The 
resulting wavelength (\ie the distance between two consecutive maxima or minima
of the vertical deviations along the arm) is 2.4 kpc, and is about twice the 
value of the separation between the complexes delineated by Cepheid stars 
(Berdnikov \& Efremov 1993) and young clusters (Avedisova 1989). This is also 
twice the average distance of the ``shingle-like" structures delineated, with 
the aid of wide-angle photographs and photoelectric surface photometry, by 
star-forming regions in this spiral arm, 1.2 kpc (Schmidt-Kaler \& Schlosser 
(1973). This structuring was derived almost three decades ago with lower
quality data (and, as correctly stated by an anonymous referee, it could now be 
repeated with much better optical, infrared, radio, and extinction data), but it
is interesting to notice that it is consistent with more recent results obtained 
with other tracers.

In the case of the gas, there are determinations of the azimuthal corrugation
scales for all known Galactic arms (Spicker \& Feitzinger 1986). This seems to
indicate that they are common features of spiral arms, but this is a difficult
study and the resulting wavelength values are less accurate than those 
estimated for stellar clusters. Moreover, the derived wavelengths depend on the
tracer used, and there are three main values reported for the Sagittarius arm: 
the shorter value (derived from OB groups and HII regions) is between 1 and 2 
kpc, the second one is between 4 and 8 kpc, and the large scale value (obtained
from HI data in the interval $l=30^{\circ}- 60^{\circ}$) has an upper limit of 
13.6 kpc (Schmidt-Kaler \& Schlosser 1973; Spicker \& Feitzinger 1986). This 
variety of different scale lengths may be indicative that, aside from possible 
errors in several of these estimates, a number of different wavelengths could 
be coexisting in the arms, or that the corrugation length varies with 
galactocentric distance. Due to obvious extinction problems, the larger scales
cannot be confirmed with stellar data, but the shorter length is about a half 
of the wavelength derived by Alfaro \etal for open clusters (again, more 
sensitive HI and CO data are now available, and interferometric HI data will 
soon be available, that could be used in a new analysis of gas corrugations 
along spiral arms).

Summarizing these results, the present observational evidence indicates that
azimuthal corrugations are probably common features in spiral arms. The 
spacings between clouds, or young stellar groups, are probably associated with 
these corrugation wavelengths. In particular, for the optical segment of the 
Carina-Sagittarius arm, the corrugation length is 2.4 kpc. This value is 
coincidental with the wavelength derived in this study for the fastest growing 
mode of the undular branch of the Parker instability.

\subsection{Superclouds}
 
In a closely related type of study, Efremov (1998) made an analysis of the
spacings between ``HI superclouds'' in Carina-Sagittarius. These are large
atomic hydrogen clouds, with masses between $10^6$ and $4\times 10^7$
$M_{\odot}$ and average particle densities between 3 and 20 cm$^{-3}$. They 
seem to be gravitationally bounded and contain giant molecular complexes, and
their molecular fraction decreases with increasing galactocentric distance. 
Also, they are found in spiral arms (Elmegreen \& Elmegreen 1987), but there
are are at least a couple of large clouds in our Galaxy that may belong to this
category and are not clearly inside a spiral arm (G216-2.5 in the outer Galaxy, 
and G2817+0.05 in the inner Galaxy. Their masses are close to $10^6$ 
$M_{\odot}$, the lower mass limit for superclouds; see Maddalena \& Thaddeus 
1985; Williams \& Maddalena 1996; Minter \etal 2001)). 

Efremov found a bimodal distribution in the spacing of these clouds, and the
two peaks are centered around the values $0.1R_{0}$ and $0.2R_{0}$. He noted 
that this type of distribution also appears in other spirals such as NGC 1365, 
NGC 2395 and NGC 613 (Elmegreen \& Elmegreen 1983). The shorter scales seem to 
be associated with inner galactic regions, while the outer segments of the arms
show larger average separations. This is more evident along the segment in the 
first galactic quadrant, where the HI superclouds are separated by shorter 
distances (Elmegreen \& Elmegreen 1987). A similar study in a sample of spiral
galaxies by Elmegreen \& Elmegreen (1983), shows that the spacings of giant HII
regions and superclouds depends on the size of the host galaxy: the typical 
distance ($L$) between adjacent star-forming regions scales linearly with the 
photometric radius ($R_{25}$) of the galaxy, $L\sim 0.23 R_{25}$. Taking 
$R_{25}=10$ kpc for our Galaxy (de Vaucoleurs 1979), the corresponding mean 
distance between giant HII regions along the arms is 2.3 kpc. This value agrees
with the second peak value found by Efremov and is very close to the 
corrugation length found by Alfaro \etal (1992a). This latter coincidence may 
be only fortuitous, but the spacings and masses of superclouds are certainly
similar to those obtained in our model.

\subsection{Magnetic Fields}

The ordered component of the magnetic field in the disk of the Milky Way, and 
external spirals, have both radial and azimuthal components (see Beck \etal 
1996; Valle 1998; Beck 2001). Many details of the field structure are still 
poorly known, but the field of the disk extends to high-galactic latitudes and 
displays several reversals both inside and outside the Solar circle. The 
strength of the local regular field is between 3 and 5 $\mu$G (the total field
strength near the Sun is 4 -- 8 $\mu$G, and this value increases by about 50\% 
at 3 kpc from the Galactic center). The azimuthal field pattern runs nearly 
parallel to the spiral arms, but the pitch angles of the "magnetic arms" are 
probably smaller than those derived for the arms in gas and stars. Actually, 
the stellar and gaseous arms observed in infrared and optical pictures of 
external galaxies do not always coincide, and in some cases they are very 
different (for instance, NGC 309 is a multi-arm spiral in the optical, but is a
barred spiral with two-arms in the infrared; see chapter 5 of Bertin \& Lin 
1996). Some of these differences are also apparent in the magnetic arms of a 
number of external galaxies, where the field maxima do not coincide with the 
locations of the optical arms. Despite these differences, however, the ordered
fields are always aligned almost parallel to the spiral arms, and can reach
average strength values of up to 20 $\mu$G (see Beck \etal 1996).

In the case of M31, the magnetic field shows a ring-like structure at a radius
of about 10 kpc from the center, that follows the CO and dust emission from the
arms (see Beck 2000). The alignment is very clear in the whole ring and
suggests that the $B$-field is anchored in the gaseous clouds of the arms. The 
regular field, then, runs along the arm but it also shows systematic 
fluctuations around the mean orientation (Beck \etal 1989). These fluctuations,
which have a conspicuous arc-like pattern, indicate azimuthal corrugations in 
the magnetic field along the arm, with a scale length of about 4.5 kpc. Thus,
the corrugation pattern is also apparent in the $B$-fields of the 10 kpc ring 
in M31, and the observed arc-like structures are similar to those obtained in
our model (see Figure 4d).

\section{Discussion}

Here we have presented 3-D models of the Parker instability including the
presence of a spiral wave. We assume a magnetized, isothermal thick disk
model that is Parker unstable, with parameters appropriate for the optical
segment of the Carina-Sagittarius arm. The spiral wave is treated as a
perturbation in the gravitational field, and the initial $B$-field lines
are oriented parallel to the spiral arm. The wavelength of the resulting
fastest growing mode along the arm is about 2.4 kpc, very similar to the
corrugation distance derived by Alfaro \etal (1992a). The corresponding
growth time scales are about 3$\times 10^7$ year. These values are only 20
\% smaller than those derived for the Solar circle (Paper I), but the
most relevant outcome of this study is the role played by the instability
in structuring the spiral arm.

The principal results of the 3-D models are: $(i)$ The presence of the
spiral wave effectively quenches the development of the interchange mode.
Thus, the regions inside the arm are dominated by the undular mode, while
the inter-arm regions are subject to the mixed mode. $(ii)$ As the undular
mode grows along the arm, mass gathering in the magnetic valleys creates
large gas condensations. $(iii)$ The final result of the instability is
an antisymmetric distribution of large clouds along the arm, and
smaller scale interstellar structures in the inter-arm regions. $(iv)$ The
$z$-component of the velocity field along the arm is also undulated, with
an amplitude of about 10 km/s.

The main gas condensations, with masses within $10^6$ -- $10^7$ $M_{\odot}$,
have properties similar to those of the HI superclouds described by Elmegreen
\& Elmegreen (1987). They are in the correct mass range and, also, are
formed from the diffuse HI medium located along the spiral arms. Our model
indicates that, regardless of the presence of large scale gravitational
instabilities in sheared magnetized regions (as have been discussed by 
Elmegreen 1987), these HI superclouds can be formed by the undular mode 
inside the arm. A logical consequence of this mechanism is that the resulting
HI superclouds are mainly located along spiral arms. These clouds are natural
seeds for the subsequent formation of giant molecular complexes, which can be
formed in the central cloud regions, and are arranged in an antisymmetric 
fashion with respect to the midplane. We are currently performing 2-D 
simulations including the formation of molecular hydrogen, that indicate that
this is certainly the case. The spacings of these clouds and their corresponding 
star-forming regions, then, should be a half of the wavelength of the fastest 
growing undular mode. For the conditions of the optical segment of the 
Carina-Sagittarius arm, this is about 1.2 kpc, very similar to the observed 
spacings of gaseous features and HII regions. This also gives support to the 
idea presented by Beck \etal (1989) that the corrugations of the $B$-field in 
M31, with length scales of about 4.5 kpc, may be due to the Parker-Jeans 
instability. One obvious corollary is that, given that the wavelength of the 
most unstable perturbation depends on the disk parameters, the spacings and 
corrugation lengths should vary with galactocentric distance.

Self-gravity, differential rotation, randomly oriented $B$-fields, cosmic rays,
and the gas flow through the arm were not included in this first work. These 
effects are important but they are difficult to include, with our present 
resources and in a self-consistent manner, in a thick gaseous disk with a 
spiral wave. Some hints from previous studies, however, can be used to 
illustrate their possible influence. For instance, the spiral wave rotates as a
rigid body and the shear due to differential rotation will affect mainly the 
inter-arm regions. In addition, the 3--D results with rotation discussed by 
Kim, Ryu \& Jones (2001) indicate that the inter-arm structures will be more 
affected and may become filamentary. Cosmic rays, on the other hand, will 
supply buoyancy to the magnetic tubes, reducing the time scales for the growth 
of the undular mode (\eg Hanasz \& Lesch 2000). Another effect is  that raising
magnetic loops will be twisted by differential rotation, leading to 
reconnection and dynamo action at a certain height. Also, as discussed by 
Elmegreen (1987) for gravitational instabilities, the magnetic field transfers
angular momentum in a sheared disk and reduces the effects of differential 
rotation. These issues require further higher resolution studies including the
effects of randomly oriented $B$-fields, that also modify the growth and 
wavelengths of the interchange mode (a random component strongly reduces the
growth of this mode; Kim \& Ryu 2001). Self-gravity, on the other hand,
increases the growth rate of the undular instability but the combined 
Jeans-Parker instability does not change the wavelength of the most unstable 
mode (Elmegreen 1982a,b; Chow \etal 2000; Lee \& Hong 2001). This will speed 
up the process, and will also lead to more compact and denser condensations 
inside the arm. Another very interesting and important effect discussed by
Masset \& Tagger (1996, 1997) is the possible exchange of waves between the 
gaseous and stellar disks that, aside from exciting warps, can also lead to 
corrugations. The inclusion of these effects, then, will tend to increase 
the contrast between arm and inter-arm regions, but will not change the basic 
conclusion of this study. Further observational and theoretical studies 
addressing these issues will shed more light on these important questions.
In particular, forthcoming high-resolution studies of the atomic and molecular
gas components of our Galaxy, such as the BU-FCRAO Galactic Ring Survey
(\ie Simon \etal 2001; Jackson \etal 2001), will provide maps of the 
Carina-Sagittarius arm in great detail. 

\acknowledgments
It is a big pleasure to thank Don Cox, Marco Martos, Fr\'ed\'eric Masset, 
Alfredo Santill\'an, and Michel Tagger for stimulating and informative 
discussions during the development of this project. We thank Rainer Beck for
information about the $B$-fields in spirals. We also thank the comments and 
questions made by the referee, that helped us to clarify the presentation.
We are grateful to the Guillermo Haro Program of INAOE for hosting a 
magnificent workshop on the Dynamics of Disk Galaxies, where part of the 
present study was performed. This work has been partially supported by 
bilateral agreements between CONACYT-Mexico with KOSEF-Korea and CSIC-Spain. 
JF thanks the Instituto de Astrof\'{\i}sica de Andaluc\'{\i}a and Seoul 
National University for their warm hospitality, and acknowledges partial 
support by DGAPA-UNAM grant IN118401 
and by a R\&D CRAY Research grant. JK also acknowledges the warm hospitality 
of the Instituto de Astronom\'{\i}a-UNAM. The work of SSH was supported in 
part by a grant from the Korea Research Foundation made in the year 1997. EJA 
acknowledges the hospitality of the Instituto de Astronom\'{\i}a-UNAM, and 
partial support from DGICYT through grants PB97-1438-C02-02 and by the 
Research and Education Council of the Autonomous Government of Andaluc\'{\i}a 
(Spain).

\clearpage

\centerline{\bf Figure Captions}

\figurenum{1}
\figcaption{
The run of the rms velocities as a function of time. The natural log is
used along the ordinate. The solid
line has a slope equal to the growth rate for the fastest growing undular
mode.  The units of velocity and time are the isothermal sound speed,
8.4 km sec$^{-1}$, and sound crossing time over the effective
scale height, $1.5 \times 10^{7}$ yr, respectively.  
}

\figurenum{2}
\figcaption{
Perspective views of density structures, velocity vectors, and
magnetic field lines at three selected times: $t$ = 15, 25, and 32. 
Again, the time unit is $1.5 \times 10^{7}$ yr. The computational box 
(4 kpc)$^3$ is oriented in such a way that the radial ($x$), azimuthal 
($y$), and vertical ($z$) directions are from left to right, from near to 
far, and from bottom to top, respectively. Color-coded densities are plotted
in the central $(x,z)$ plane. Colors are mapped
from red to purple as density decreases. The velocity vectors are
represented by red arrows, and the velocity unit is rescaled at each time.
The maximum velocity vectors in the three frames are 6.47, 49.73, and 41.33 
km sec$^{-1}$, respectively.  Two isodensity surfaces, with the reference 
density $n=0.04$ cm$^{-3}$, are also shown above and below the midplane.
Thirty-two field lines, whose starting points lie along a vertical line 
in the front $(x,z)$ plane, are plotted as blue tubes.
}

\figurenum{3}
\figcaption{
Column density maps, integrated along the $z$-axis from -2 kpc to +2 kpc,
for the same selected times displayed in Fig. 2. The column density values
are normalized to the initial value, $1.1 \times 10^{21}$ cm$^{-2}$.  Thick 
solid lines with this initial value separate the under-dense and over-dense
regions.  The thin solid (dotted) lines show normalized column densities 
larger (smaller) than 1.0, and the interval between successive lines is 0.1. 
A gray scale image is also overlaid, to provide a better
representation of the over-dense and under-dense regions.
The time and length units are $1.5 \times 10^{7}$ years and kpc,
respectively.
}

\figurenum{4}
\figcaption{
Sliced density maps at $t=32$. (a) The left panel represents a gray 
image densities on the $(x,y)$ plane, together with isodensity contours.
Density levels are from 0.4 to 1.2 with 0.2 interval.  A cubic-polynomial
fitting curve that passes through the peaks of the main condensations
along the spiral arm is also included.  (b) The right panel 
represents a gray image of the vertical density distribution along the 
curve of the left panel superposed on isodensity contours. 
The units of time, length and density are $1.5 \times 10^{7}$ yr,
kpc, and 1.4 cm$^{-3}$ (a midplane value), respectively.
}

\figurenum{5}
\figcaption{
Magnetic field lines near the central (a) $(y,z)$ and (b) $(x,y)$ planes,
together with isodensity surfaces with $n=0.8$ cm$^{-3}$ at $t=32$.   
Again, the time unit is $1.5 \times 10^{7}$ yr.
}

\figurenum{6}
\figcaption{
Position-velocity map along the spiral arm. This figure shows an important
kinematic signature of the process, The positions run along the curve defined
in Figure 4a, and the velocity corresponds to the $z$-component at $t=32$. 
The time unit is $1.5 \times 10^{7}$ yr.
}
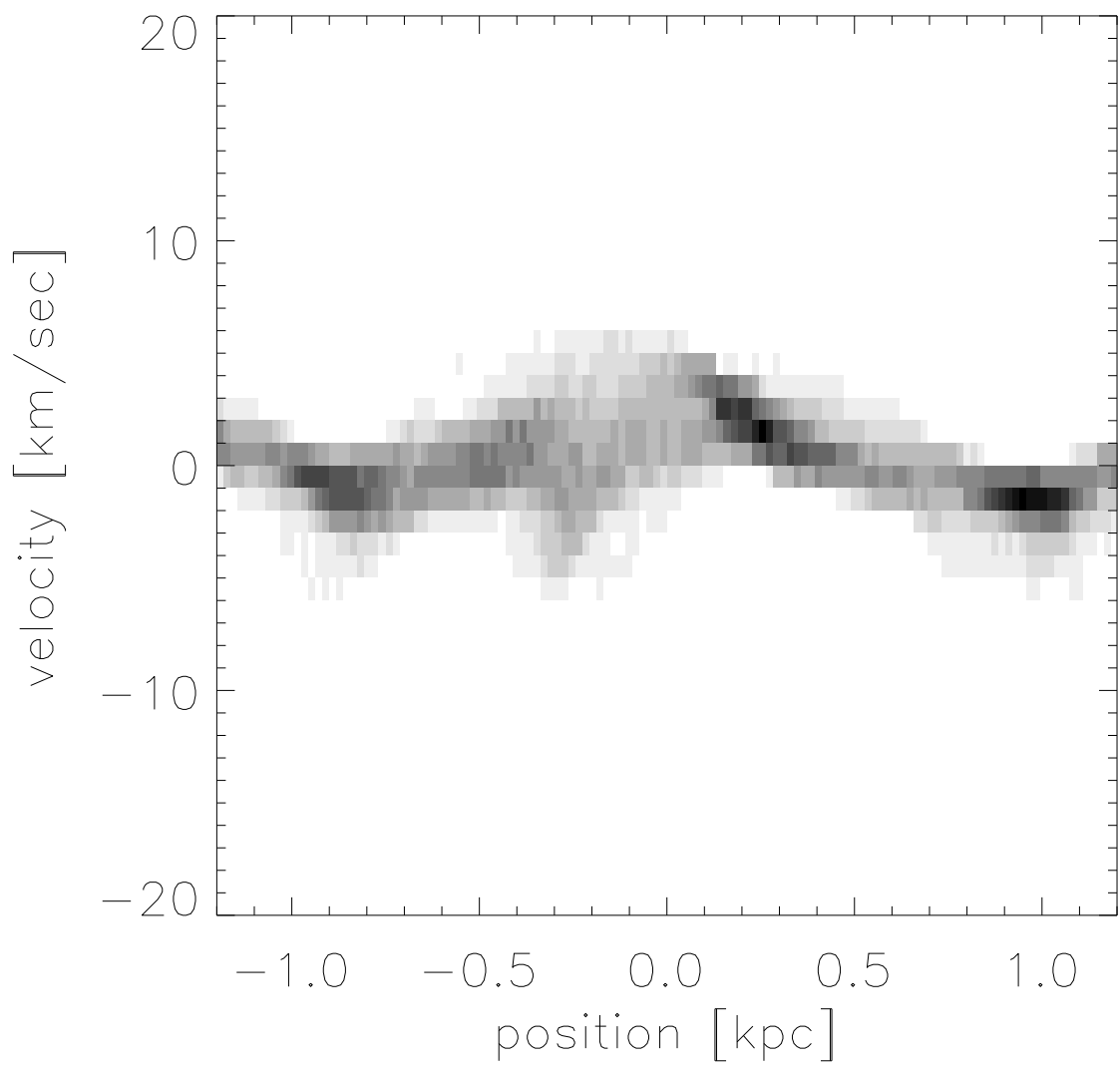
\end{document}